\documentstyle[aps]{revtex}
\begin{document}
\voffset 0.8truecm
\title{
Quantum Cloning Machines of a $d$-level System}

\author{
Heng Fan$^{a}$, Keiji Matsumoto$^a$, Miki Wadati$^b$}
\address{
$^a$Imai quantum computing and information project,
ERATO, \\
Japan Science and Technology Corporation,\\
Daini Hongo White Bldg.201, Hongo 5-28-3, Bunkyo-ku, Tokyo 133-0033, Japan.\\
$^b$Department of Physics, Graduate School of Science,\\
University of Tokyo, Hongo 7-3-1, Bunkyo-ku, Tokyo 113-0033, Japan.
}
\maketitle
                                                        
\begin{abstract}
The optimal $N$ to $M$ ($M>N$) quantum cloning machines for
the $d$-level system are presented. The unitary cloning transformations
achieve the bound of the fidelity.          
\end{abstract}
       
\baselineskip 0.5truecm

No-cloning theorem is one of the most fundamental differences
between classical and quantum information theories.
It tells us that an unknown quantum state can not be 
copied perfectly\cite{WZ}. This no-cloning theorem has
important consequences for the whole quantum information
processing\cite{NC}. However, no-cloning
theorem does not forbid imperfect cloning, and it is interesting
to know how well we can copy an unknown quantum state.
Bu\v{z}ek and Hillery\cite{BH} introduced
a universal quantum cloning machine (UQCM) for an arbitrary
pure input state. It produces two identical copies whose quality
is independent of the input state. It was proved that
the Bu\v{z}ek and Hillery UQCM is optimal by
using the fidelity as measurement \cite{BDEF,BEM,GM,W}.
The general optimal quantum cloning transformation with $N$
identical pure input states and $M$ copies
was proposed by Gisin and Massar\cite{GM,BEM}. And the no-cloning
theorem was also extended to other cases\cite{BCFJ,KI}.

The UQCM presented by Bu\v{z}ek, Hillery and Gisin, Massar
is for the 2-level quantum system\cite{BH,GM}. The 2-dimensional Hilbert
space is spanned by 2 orthonormal basis vectors $|1\rangle ,
|2\rangle $ ($|\uparrow \rangle ,|\downarrow \rangle $).
For a $d$-level quantum system, the $N$ to $M$ optimal quantum
cloning is formulated by Werner\cite{W} and Keyl and Werner\cite{KW}.
They have shown that the optimal cloning map to obtain
$M$ optimal clones from $N$ identical input states is the
projection of the direct product of the $N$ input states and
$M-N$ identity states onto the symmetric subspace of $M$
particles.
The optimal fidelity
is obtained from the so-called Black Cow factor.
Bu\v{z}ek and Hillery \cite{BH1} presented the universal 1 to 2
quantum cloning transformation of states in
$d$-dimensional Hilbert space. Albeverio and Fei\cite{AF} extended
this result to 1 to $M$ cloning, and a special case of
$N$ to $M$ cloning in which the input state
is a restricted $N$ identical $d$-level quantum system.
In this paper, generalizing the results in Ref.\cite{BH1,AF},
we shall present
the optimal $N$ to $M$ unitary cloning transformation for the $d$-level
system.
The fidelity achieves the optimal fidelity
given by Werner \cite{W}.
Our results recover
the previous results in Ref.\cite{BH,GM,BH1,AF} for special values of
$N$, $M$ and $d$.
The cloning transformation presented in
this paper should be the physical implementation of the
optimal cloning map given in Ref.\cite{W,KW}. The result
is useful in obtaining the quantum networks \cite{BBHB}
and the remote information concentration\cite{MV}.

A $d$-level quantum system is spanned by the orthonormal basis
$|i\rangle $ with $i=1, \cdots ,d$. And an arbitrary pure
state takes the form $|\Psi \rangle =\sum _{i=1}^d x_i|i\rangle $ with
$\sum _{i=1}^d|x_i|^2=1$. The $N$ to $M$ quantum cloning means that
we map by unitary transformation $N$ pure input states
\begin{eqnarray}
|\Psi \rangle ^{\otimes N}\otimes R\equiv
\sum _{{\bf n}=0}^N \sqrt {\frac {N!}{n_1!\cdots n_d!}}
x_1^{n_1}\cdots n_d^{n_d}|{\bf n}\rangle \otimes R
\label{dinput}
\end{eqnarray}
to $M$ copies in $|{\bf m}\rangle \otimes R_{\bf nm}$,
where vector ${\bf n }$ denotes $n_1, \cdots , n_d$,
$|{\bf n}\rangle =|n_1, \cdots n_d\rangle $ is
a completely symmetric and normalized state with
$n_i$ systems in $|i\rangle $, this state is invariant
under permutations of all $N$ $d$-level qubits.
$R$ denotes $M-N$ blank
copies and the initial state of the cloning machine,
$R_{\bf nm}$ are internal
states of the cloning machine, where $\sum _{{\bf n}=0}^N$
means sum over all variables under the condition
$\sum_{i=1}^d n_i=N$,
we also have $\sum_{i=1}^d m_i=M$.
More explicit, we need to find unitary transformations
$U_{NM}$ to map $|{\bf n}\rangle $ to $|{\bf m}\rangle $.
We use fidelity $F$ to describe the quality of the copies
$F=\langle \Psi |\rho ^{out}|\Psi \rangle $,
where $\rho ^{out} $ denotes reduced density operator of each
output $d$-level qubit by taking partial trace over all
but one output $d$-level qubits.

We propose the $N$ to $M$ quantum cloning transformation for
$d$-level quantum system as follows,
\begin{eqnarray}
U_{NM}|{\bf n}\rangle \otimes R=
\sum _{{\bf j}=0}^{M-N}\alpha _{\bf nj}
|{\bf n}+{\bf j}\rangle \otimes R_{{\bf j}},
\label{result1}
\end{eqnarray}
where ${\bf n}+{\bf j}={\bf m}, i.e.,
\sum _{k=1}^dj_{k}=M-N$, $R_{\bf j}$ denotes the orthogonal
normalized internal states of the UQCM, and
\begin{eqnarray}
\alpha _{\bf nj}=
\sqrt {\frac {(M-N)!(N+d-1)!}{(M+d-1)!}}
\sqrt {\prod _{k=1}^d\frac {(n_k+j_k)!}{n_k!j_k!}}.
\label{result2}
\end{eqnarray}
This explicit construction
of the unitary cloning transformation is the essence of this paper.
The dimension of the ancilla is $\frac {(M-N+d-1)!}{(M-N)!(d-1)!}$.
The $N$ to $M$ cloning
in Ref.\cite{AF} corresponds to the case $n_i=N$ in the above
relations.

As an example, we present the $N$ to $M$ quantum cloning transformation
for the 2-level system since it is simple and is
the physically interesting case.
Gisin and Massar's UQCM from $N$ to $M$
for the 2-level system is well known\cite{GM}.
Our quantum cloning transformation turns out to be
a different but an equivalent form.
Suppose the pure input state takes the form $|\Psi \rangle =
\alpha |\uparrow \rangle +\beta |\downarrow \rangle $,
the $N$ identical input qubits are thus written as
\begin{eqnarray}
|\Psi \rangle ^{\otimes N}=\sum _{k=0}^N\alpha ^{N-k}\beta ^k
\sqrt{C_N^k}|(N-k)\uparrow ,k\downarrow \rangle ,
\end{eqnarray}
where 
$|(N-k)\uparrow ,k\downarrow \rangle $ denotes the symmetric and
normalized state with $N-k$ qubits in the state $|\uparrow \rangle $ and
$k$ qubits in the state $|\downarrow \rangle $,
and we have $C_N^k=\frac {N!}{(N-k)!k!}$ in standard notation.
According to (\ref{result1},\ref{result2}), we can write the
quantum cloning transformation for the qubits case as,
\begin{eqnarray}
&&U_{NM}|(N-k)\uparrow ,k\downarrow \rangle \otimes R
=\sum _{j=0}^{M-N}
\alpha _{kj}|(M-k-j)\uparrow ,(k+j)\downarrow \rangle \otimes R_j,
\nonumber \\
&&\alpha _{kj}=\sqrt{\frac {(M-N)!(N+1)!}{k!(N-k)!(M+1)!}}
\sqrt{\frac {(k+j)!(M-k-j)!}{j!(M-N-j)!}}.
\label{2d}
\end{eqnarray}
The reduced density operator describing the state of each ouput is
given by
\begin{eqnarray}
\rho ^{out}&=&|\uparrow \rangle \langle \uparrow |
\left( \sum _{j=0}^{M-N}\sum _{k=0}^N
|\alpha |^{2N-2k}|\beta |^{2k}\alpha _{kj}^2C_N^k\frac {M-k-j}{M}
\right) \nonumber \\
&+&|\uparrow \rangle \langle \downarrow |
\left( \sum _{j=0}^{M-N}\sum _{k=0}^{N-1}
|\alpha |^{2N-2k}|\beta |^{2k}
\frac {\beta ^*}{\alpha ^*}
\alpha _{kj}\alpha _{(k+1)j}\sqrt {C_N^kC_N^{k+1}}
\frac {\sqrt{(k+j+1)(M-k-j)}}{M}\right)
\nonumber \\
&+&|\downarrow \rangle \langle \uparrow |
\left( \sum _{j=0}^{M-N}\sum _{k=0}^{N-1}
|\alpha |^{2N-2k}|\beta |^{2k}
\frac {\beta }{\alpha }
\alpha _{kj}\alpha _{(k+1)j}\sqrt {C_N^kC_N^{k+1}}
\frac {\sqrt{(k+j+1)(M-k-j)}}{M}\right)
\nonumber \\
&+&|\downarrow \rangle \langle \downarrow |
\left( \sum _{j=0}^{M-N}\sum _{k=0}^N
|\alpha |^{2N-2k}|\beta |^{2k}\alpha _{kj}^2C_N^k\frac {k+j}{M}
\right). 
\end{eqnarray}
With the help of the definition,
the fidelity of the quantum cloning transformation can thus be
calculated as
\begin{eqnarray}
F=\frac {MN+M+N}{M(N+2)}.
\end{eqnarray}
This is the optimal fidelity which can be achieved\cite{GM,BEM}.
We thus see that the quantum cloning transformation (\ref{2d}) is optimal
and is equivalent to the UQCM given
by Gisin and Massar\cite{GM}.

Finally, we present the formula for the $d$-level quantum system.
The input states are $N$ identical states given in (\ref{dinput}).
By use of the transformation in (\ref{result1},\ref{result2}),
the output reduced density operator at each $d$-level qubit can
be written as
\begin{eqnarray}
\rho ^{out}&=&\sum _{i=1}^d|i\rangle \langle i|
\left( \sum _{{\bf n}=0}^N\sum _{{\bf j}=0}^{M-N}
(\prod _{k=1}^d\frac {|x_k|^{2n_k}(n_k+j_k)!}{(n_k!)^2j_k!})
\frac {N!(M-N)!(N+d-1)!}{(M+d-1)!}\frac {(n_i+j_i)}{M}\right)
\nonumber \\
&+&\sum _{i\not= l}^d|i\rangle \langle l|
\left( \sum _{{\bf n}=0,n_l<N}^N\sum _{{\bf j}=0}^{M-N}
(\prod _{k=1}^d\frac {|x_k|^{2n_k}(n_k+j_k)!}{(n_k!)^2j_k!})
\frac {x_l^*}{x_i^*}
\frac {N!(M-N)!(N+d-1)!}{(M+d-1)!M}\frac {n_i(n_l+j_l+1)}{n_l+1}\right).
\nonumber \\
\end{eqnarray}
The fidelity can be calculated as
\begin{eqnarray}
F=\frac {N(d+M)+M-N}{(d+N)M}.
\label{gfid}
\end{eqnarray}
This is the optimal fidelity achieved by universal
quantum cloning transformation for a $d$-level system given by
Werner\cite{W} and Keyl and Werner\cite{KW}. Thus the UQCM given in
(\ref{result1},\ref{result2}) is optimal.
As the case of 1 to 2 UQCM \cite{BH1},
the cloning transformation in this letter can also be applied
for the arbitrary impure states cloning by considering
the input density operator as
$\sum _{ij}A_{ij}|\Psi _i\rangle ^{\otimes N~N\otimes}\langle \Psi _j|$.
The fidelity keeps the same (\ref{gfid}) as the case
of pure input state.

To summarize, we have presented the 
optimal unitary cloning transformations with $N$ identical
unknown input states to $M$ copies for the $d$-level system.
As the results in Ref.\cite{BH,GM,W,KW,BH1,AF}, the output of
the UQCM in this letter is completely symmetric. 
If we want to have a higher fidelity, for example, in the first
output $d$-level qubit than the optimal fidelity, the fidelity
in other output $d$-level qubits will decrease. We give a
simple and limited example: the input are $N$ identical states
$|\Psi \rangle |\Psi \rangle ^{\otimes N-1}$, in the
cloning transformation, we keep the first $d$-level qubit unchanged while
the other $N$-1 qubits are changed by the optimal cloning transformation.
The fidelity for the first output $d$-level qubit is 1, while the
fidelity of other $d$-level qubits equal to the case of $N-1$
to $M-1$ cloning
whose fidelity is smaller than the case of $N$ to $M$ cloning.
In the optimal UQCM (\ref{result1}), for a fixed $M$,
because that the input $|{\bf n}\rangle $ always keeps in
the output state $|{\bf m}\rangle $,
we can see that the more quantum information
is available, the better fidelity of the copies we can get.
We remark that, similar to 2-dimensional case\cite{GM},
the $N$ to $N+1$ UQCM is much simple, the right hand side of
(\ref{result1}) contains only $d$ terms for every $|{\bf n}\rangle $,
and the UQCM needs only $d$ internal states altogether.

In order to construct the quantum networks to realize the optimal
quantum cloning, the results in this paper which give the
cloning transformations for all symmetric input states should be
useful.  Different from the quantum networks, the optimal
quantum cloning realized via photon stimulated emission
was proposed for the 2-level system in \cite{SWZ,KSW}. For the 
d-level system, a similar result should also be found with a
generalized photon-atom interaction Hamiltonian.

\noindent {\bf Acknowlegements:} One of the authors, H.F.
acknowleges the support of JSPS, and the hospitality of Wadati group in
Department of Physics, University of Tokyo where part of
this work were done.
We thank X.B.Wang and G.Weihs
for useful discussions.

\end{document}